\documentclass[sigconf,10pt]{acmart}

\usepackage[english]{babel}
\usepackage{blindtext}

\renewcommand\footnotetextcopyrightpermission[1]{} 
\setcopyright{none}

\acmDOI{}

\acmISBN{}

\acmConference[]{}
\acmYear{2022}

\acmPrice{}

\usepackage{xcolor}
\usepackage{amsfonts}
\PassOptionsToPackage{hyphens}{url}\usepackage{hyperref}
\usepackage{color}
\usepackage{graphics}
\usepackage{graphicx}
\usepackage{url}
\usepackage{listings}
\usepackage{multicol}
\usepackage{multirow}
\usepackage[scaled]{helvet}
\usepackage{rotating}
\usepackage{xspace}
\usepackage[ruled,vlined]{algorithm2e}
\usepackage{comment}
\usepackage{enumitem}
\usepackage{amsmath}
\usepackage{mathrsfs}
\usepackage{cancel}
\usepackage{cleveref}
\usepackage{subcaption}

\newcommand{\name}{MIRAGE\xspace}

\begin{document}
\title{Users are Closer than they Appear: User Location Privacy against WiFi snoopers}
\title{Users are Closer than they Appear: Protecting User Location from WiFi APs}


\author{Roshan Ayyalasomayajula, Aditya Arun, Wei Sun, and Dinesh Bharadia}
\affiliation{UC San Diego}

\begin{abstract}
WiFi-based indoor localization has now matured for over a decade.
Most of the current localization algorithms rely on the WiFi access points (APs) in the enterprise network to localize the WiFi user accurately.
Thus, the WiFi user's location information could be easily snooped by an attacker listening through a compromised WiFi AP.
With indoor localization and navigation being the next step towards automation, it is important to give users the capability to defend against such attacks.
In this paper, we present \name, a system that can utilize the downlink physical layer information to create a defense against an attacker snooping on a WiFi user's location information.
\name achieves this by utilizing the beamforming capability of the transmitter that is already part of the WiFi protocols.
With this initial idea, we have demonstrated that the user can obfuscate his/her location from the WiFi AP always with no compromise to the throughput of the existing WiFi communication system, and reduce the user location accuracy of the attacker from 2.3m to more than 10m.
\end{abstract}

\maketitle


\section{Introduction}

\begin{figure}
    \centering
    \includegraphics[width=0.8\linewidth]{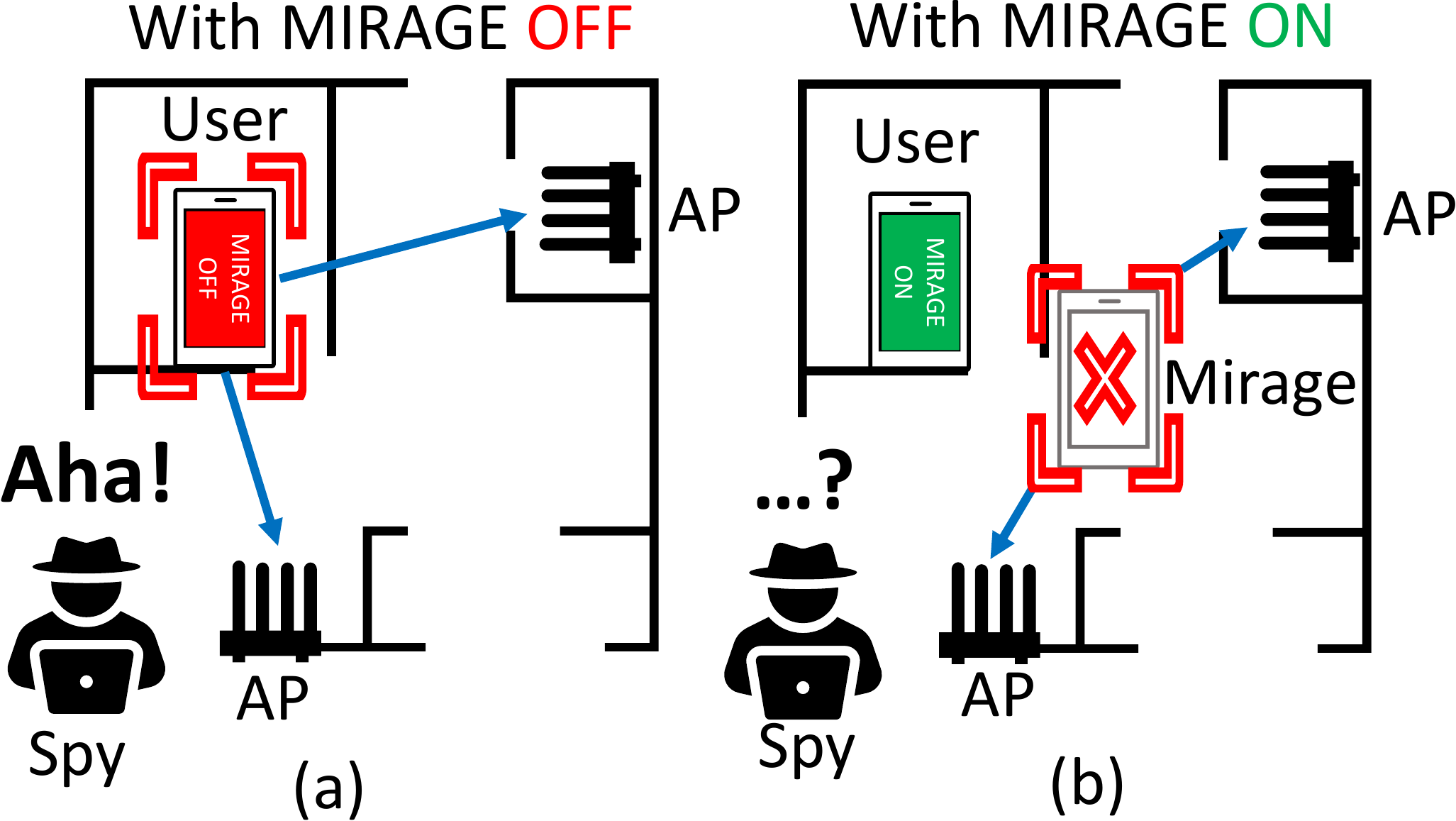}
    \caption{\name: (a) Shows the typical direction finding (e.g., AoA) based indoor device localization. (b) Shows the obfuscation that \name provides, which enables the users' location privacy.}
    \label{fig:mirage-intro}
\end{figure}

Proliferation of wireless sensing and localization has enabled the widely deployed Wi-Fi APs to not only provide Internet connectivity but also sense the user's location. Specifically, location-based services in indoor settings have gained interest, especially in the recent times for contact tracing, indoor navigation, or density monitoring. For example, Qualcomm proposes to deploy the Wi-Fi APs for joint wireless communication and sensing in the enterprise network~\cite{qualcomm} and the upcoming 5G deployments also claim to provide location services~\cite{alliance20155g}. However, this would lead to potential breaches of Wi-Fi users' private location information and consequently other sensitive information (Fig.~\ref{fig:mirage-intro}(a)). 
A simple example could be enterprise WiFi network deployed in the malls~\cite{malls_loc} already try to get accurate user locations.
This location data collected in the malls can be used to stalk users, track a user's interactions with other users or even analyze their spending trends to infer private information (e.g., sex, age, personal preferences).
We present \name, an algorithm that the user can employ on their devices (e.g., smartphone) to maintain their location privacy if desired without compromising their Wi-Fi's quality of service (Fig.~\ref{fig:mirage-intro}(b)).

The typical enterprise networks estimate the user locations using a single access point (AP) to estimate the user's distance and direction.
They can estimate user's range using existing 802.11mc~\cite{martin2020ranging} protocols that can estimate the received Wi-Fi signals Time-of-Flight (ToF). Additionally, these Wi-Fi APs, typically equipped with multiple antennas, can also measure the user's direction by estimating the received signals Angle-of-Arrival (AoA) using algorithms such as SpotFi/MUSIC~\cite{spotfi}. ToF-based range estimates usually require multiple packet exchanges with the user and a simple defense is to not respond to these requests. Furthermore, Wi-Peep~\cite{wipeep} demonstrates an attack against these 802.11 range estimation systems and proposes a few solutions for defense. Unfortunately, there seems to be no defense systems that can overcome the AoA algorithms-based direction estimation systems, as these are simply receiving the user's transmitted Wi-Fi signals. 

\begin{figure*}[ht]
    \centering
    \includegraphics[width=0.9\linewidth]{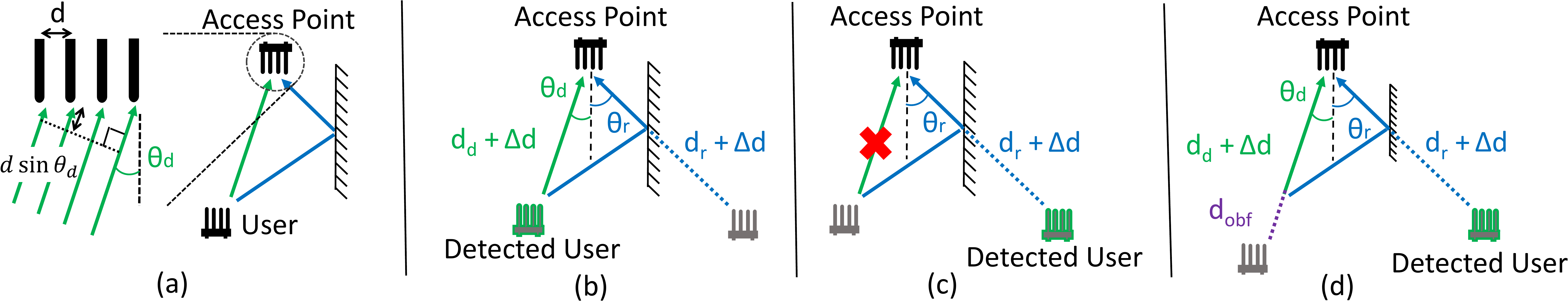}
    \caption{\name's Idea: (a) Typical indoor setting with the direct path and the first strongest reflected path. (b) AP estimates the angles of arrivals (AOAs) to be $\{\theta_d,\theta_r\}$ and relative time of flights (rToFs) as $\{d_d+\Delta d,d_r+\Delta d\}$\protect \footnotemark ($d_d < d_r$). \textit{Estimated AoA is $\theta_d$}. (c) Shows beam-nulling towards $\theta_d$. Estimated AoA ($\theta_r$) is incorrect at the cost reduced SNR. (d) \name adds delay $d_{obf}$ and makes $d_d+d_{obf} > d_r$. Estimated AoA is $\theta_r$ and no SNR reduction observed.}
    \label{fig:mirage-idea}
\end{figure*}

To understand why defending against AoA-based localization algorithms is difficult, let us understand how AoA is measured at an AP. 
The Wi-Fi signal arriving directly from a transmitter at the AP's antenna array traverses varying distances to each antenna (Fig.~\ref{fig:mirage-idea}(a)). These small distances are observable from the phase of the signal across the receive antennas. These differential phases observed across the antennas vary in accordance to the AoA and hence can be used to estimate the AoA. However, in a typical indoor environment, this Wi-Fi signal can bounce off various objects and arrive at the AP via multiple paths. These paths, dubbed multipath, can potentially corrupt the AoA of direct path. Most localization systems~\cite{ spotfi} employ a simple heuristic -- the straight-line direct path arriving at the AP travels the least distance -- and separate the multipath from the direct path by measuring the relative time of flights. Consequently, obfuscating AoA is challenging as the signal accumulates phases by physically interacting with the environment and this phase can readily be measured by AP's listening to the signal.

However, we develop a defense against this snooping attack via \name, and hence allow the user to protect their location by obfuscating primarily the direction of the user's location. The key idea is that users' direction or angle of arrival (AoA) information measured to perform localization is obfuscated by creating a `mirage' such that the users' actual direction is along one of the reflected paths, i.e. making the reflected path look like the most direct path from the user to the AP. \name achieves this obfuscation without reducing the communication data rate, i.e. it continues to use the direct path for communication. Finally, \name's mechanism for spoofing the AoA and creating a mirage also obscures the direct path's ToF information and protects against Chronos~\cite{chronos} and other ToF-based algorithms.

A naive way to create this mirage is to just simply beamform the signal transmitted by the user such that there is a null towards the access point and remove the direct path (Fig.~\ref{fig:mirage-idea}(c)). This makes the path observed in the profile correspond to the reflected path, which would obfuscate the attacker. Unfortunately, this beamforming comes at the cost of reduced SNR and hence the throughput of the network. Alternatively, in \name we develop a novel algorithm that enables us to add delay only to the direct path such that the direct path appears to have travelled more distance than the reflected path in the environment as shown in Fig.~\ref{fig:mirage-idea}{d}. Employing \name hence takes away the user certainty in the AoA of the direct path and helps protect the locations of the user. And most importantly, by preserving the direct path, \name does not affect the signal's SNR and hence the throughput.

To demonstrate \name's feasibility, we have deployed it on WARP board~\cite{warp} as the user device and commercially off the shelf (COTS) available ASUS device~\cite{pizarro2021accurate} that acts as the attacking AP. With this setup and a few experiments, we have shown

\begin{itemize}[leftmargin=*]
    \item [(a)] The naive approach of nulling towards the direct path ensures that an attacker can practically never get accurate AoAs of the Wi-Fi user, but it reduces the SNR by $6$dB.
    \item [(b)] Meanwhile, \name's design ensures obfuscation of the user's AoA by $46^\circ$  on an average creating a localization errors for AoA based systems to up to $10$~m on an average, while observing no degradation in SNR. 
    \item [(c)] \name follows Wi-Fi protocols by applying compliant precoding matrices which can be decoded by COTS APs. 
\end{itemize}

\footnotetext{ $\Delta d$ is the random distance error constant across all the paths for a given data packet, caused due to sampling frequency offsets (SFO).}

\section{Design}\label{sec:design}
In this section, we first illustrate an attack model which will violate the Wi-Fi user's location privacy. We then propose on how a straightforward idea of beam nulling to the direct path to protect the user's privacy comes at the cost of user's throughput degradation. Finally, we propose the design of \name for Wi-Fi user's location obfuscation, which makes the reflected path appear to be the direct path.

\subsection{Attack Model} 
\begin{figure}
    \centering
    \includegraphics[width=\linewidth]{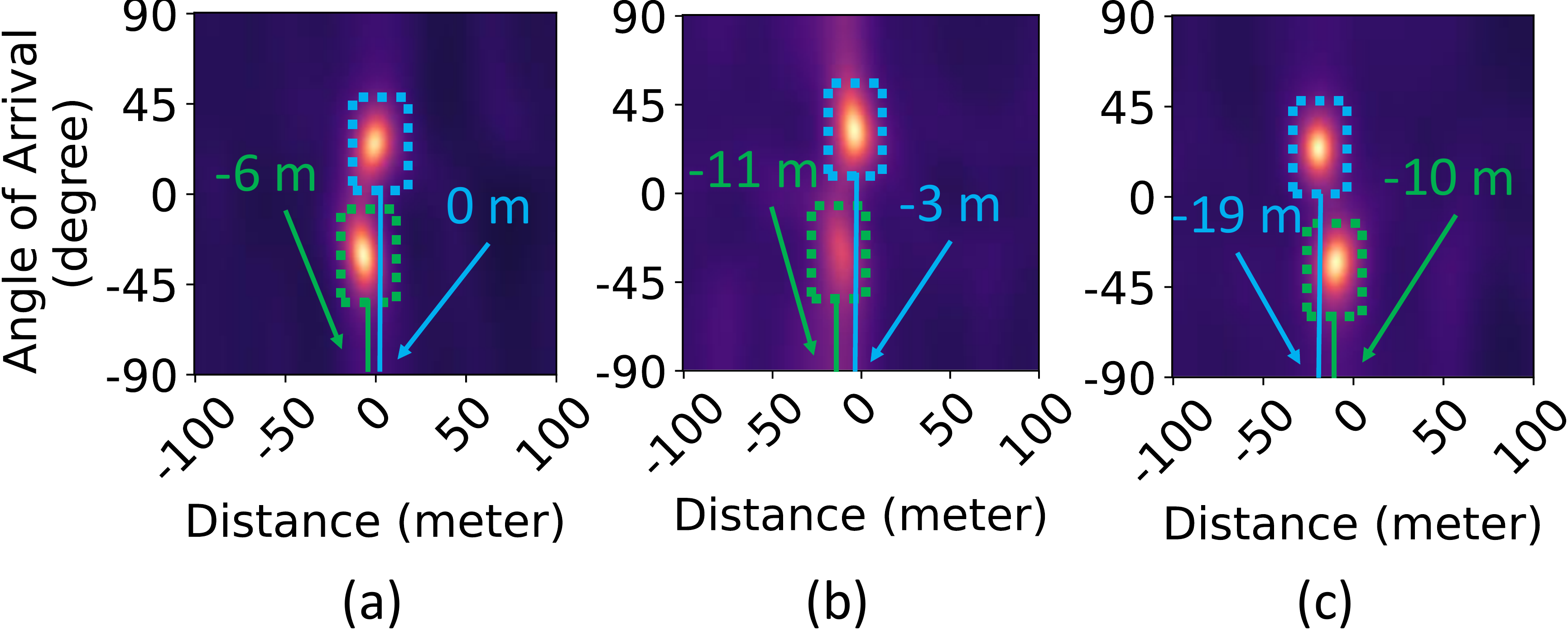}
    \caption{Angle-Distance profiles representing the direct path and reflected path angle of arrivals and their relative distance travelled. Measurement from COTS AP in (a) without \name, in (b) with nulling the direct path, in (c) with \name obfuscation applied.}
    \label{fig:profiles}
\end{figure}

The attacking Wi-Fi APs in an enterprise network (say a shopping mall) listen to the channel transmitted by the user and predict the user directions to each of the APs using state-of-the-art AoA algorithms (say SpotFi~\cite{spotfi}). The APs can use the AoAs either for
(a)`single-AP Localization' by using the ToF measurements (from Wi-Peep~\cite{wipeep} for example) for range and estimated AoA for direction, or (b) `Triangulation' by using AoAs across multiple APs to estimate user location. The AoA of the incoming Wi-Fi signal at the attacking AP can be extracted via Spotfi~\cite{spotfi} as shown in Fig.~\ref{fig:profiles}(a). As discussed, the direct path (green box) arrives earlier than the reflected path (blue box) and hence direct path's AoA can be reliably extracted even in multipath-rich environments.

\subsection{Nulling to the Direct Path}
\label{sub:nulling}
Hence, to obfuscate the AoA information at the Wi-Fi user, a straightforward idea is to null this direct path to the AP. That is the user can beamform such that there is a null in the beampattern in the direction of AP's direct path. Now the attacker Wi-Fi AP will only observe the multipath and regard it as the direct path for localization. To perform this nulling, the Wi-Fi user extracts the AoA of direct path (i.e., $\theta_d$) and AoA of the earliest reflected path (i.e., $\theta_r$) from a downlink (AP to user) channel measurement. For the sake of simplicity, we only consider two paths. To null the direct path, the Wi-Fi user needs to apply the beamforming weights $N_d(f_i,k)$ to their data streams to null in the $\theta_d$ direction, where $i$ indicates the subcarrier index and $k$ indicates the antenna index. After we `precode' the data stream $X(f_i, k)$ with nulling weights, the Wi-Fi AP will receive the following signals:
\begin{align*}
    Y(f_i,k) &=H(f_i,k)N_d(f_i, k)X(f_i,k)\\
             &=(H_d(f_i, k)+H_r(f_i,k))N_d(f_i, k)X(f_i,k)\\
             &= H_r(f_i,k)N_d(f_i, k)X(f_i,k)
\end{align*}
where $Y(f_i,k)$ and $H(f_i,k)$ indicate the received signal and wireless channel on $k$-th antenna and $f_i$ subcarrier. $H(f_i,k)=H_d(f_k, k)+H_r(f_i,k)$ where $H_d$ and $H_r$ indicate the direct path and  reflected path wireless channel. Note that $H_d(f_i, k)N_d(f_i,k)=0$ as $N_d(f_i,k)$ is chosen to lie in the null space of $H_d(f_i, k)$ \cite{li2005null}. The Wi-Fi AP will estimate the wireless channel as $H_r(f_i,k)N_d(f_i, k)$, which only contains the reflected path, thereby preventing the attacker from extracting the direct path AoA. 

This is illustrated in the angle-distance profile in Fig.~\ref{fig:profiles}(b). Note the reduced power of the direct path. However, this simple idea has two key flaws. First, nulling to the direct path will decrease the signal strength at the Wi-Fi AP and degrade the network throughput. Second, in case the nulling angle is predicted incorrectly by the user, nulling will be ineffective and a residual peak of the direct path will remain exposing the user's location. Clearly, this solution is impractical and we propose \name for user location obfuscation to eliminate this side-effect.

\subsection{Beamforming and Delaying}\label{sec:mirage-des}

To overcome this user's throughput degradation we need to ensure the direct path is preserved, as direct-path contributes to majority of the channel diversity. However the presence of the direct path will appear as a strong signal at the Wi-Fi AP and an attacker can easily extract the AoA. However, to disambiguate the effects of the inevitable reflected paths, the attacker relies on the simple heuristic that the direct path always travels the shortest path. Specifically, for a distance travelled $d$, the phases accumulated across the various subcarriers are given by $e^{-2\pi f_c \frac{d}{c}}$. By leveraging super-resolution localization algorithms~\cite{spotfi}, the attacker can hence separate the various paths in the time-domain and predict the direct path. 

Here we provide a key insight -- delaying only the direct path signal invalidates the foundational heuristic to select the correct AoA. With this in mind, we propose to beamform to direct path and multipath and add delay to the direct path. This ensures communication throughput will be preserved over the beamforming and the user's location will be obfuscated through adding the delay to the direct path. This is qualitatively illustrated in Fig.~\ref{fig:profiles}(c). The Wi-Fi user adds a delay of $15$ m in the direction of the direct path and successfully pushes the direct path peak (green box) predicted by SpotFi to the right of the multipath peak (blue box).    


\begin{figure*}[ht]
    \centering
    \begin{minipage}{0.40\linewidth}
    \includegraphics[width=\linewidth]{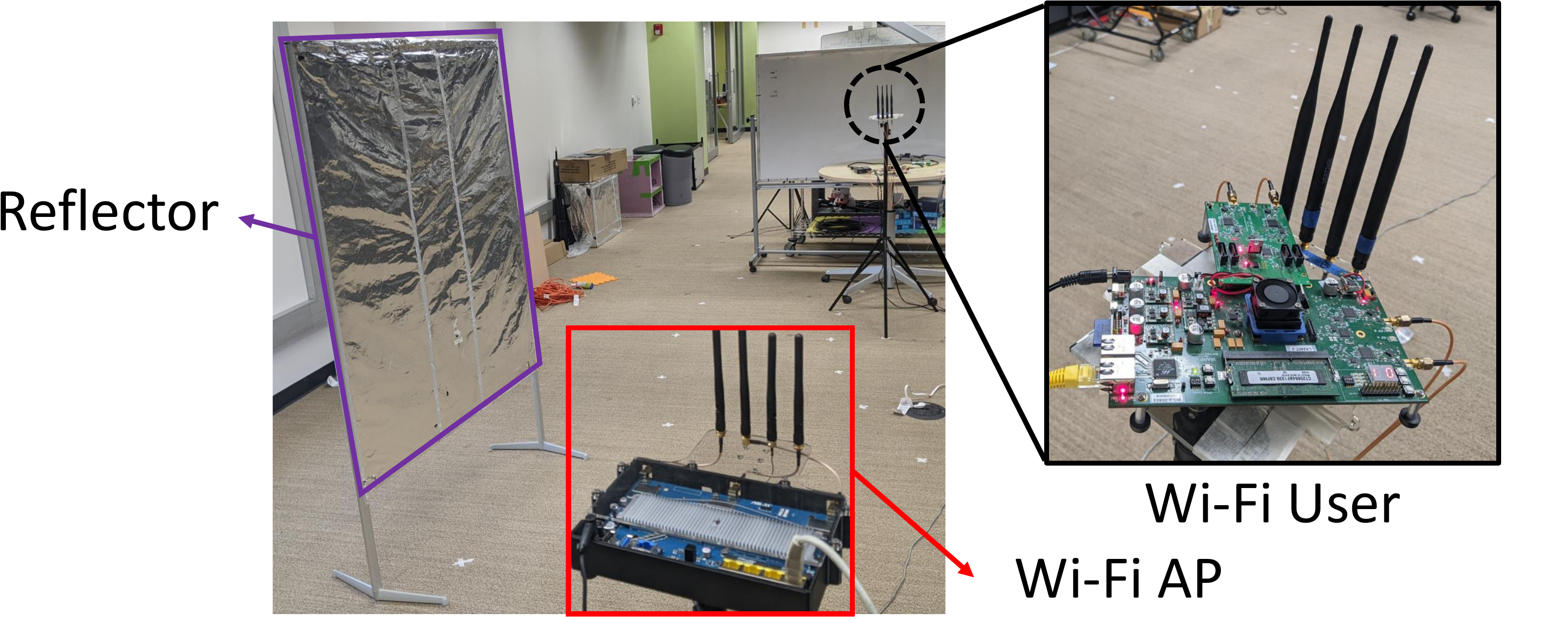}
    \subcaption{}
    \end{minipage}
    \begin{minipage}{0.49\linewidth}
        \begin{tabular}{|c|c|c|cccc|}
        \hline
        \multirow{2}{*}{} &
          \multirow{2}{*}{No obf.} &
          \multirow{2}{*}{Nulling} &
          \multicolumn{4}{c|}{\name with delay of} \\ \cline{4-7} 
         &  &  & \multicolumn{1}{c|}{0 (m)} & \multicolumn{1}{c|}{20 (m)} & \multicolumn{1}{c|}{30 (m)} & 40 (m) \\ \hline
        AoA error &
          $0^\circ$ &
          $62^\circ$ &
          \multicolumn{1}{c|}{$0^\circ$} &
          \multicolumn{1}{c|}{$58^\circ$} &
          \multicolumn{1}{c|}{$61^\circ$} &
          $53^\circ$ \\ \hline
        \begin{tabular}[c]{@{}c@{}}RSSI\\ (dBm)\end{tabular} &
          -65 &
          -71 &
          \multicolumn{1}{c|}{-64} &
          \multicolumn{1}{c|}{-64} &
          \multicolumn{1}{c|}{-64} &
          -62 \\ \hline
        \end{tabular}
        \subcaption{}
        \label{table:aoa_rssi}
    \end{minipage}
    \caption{\textbf{(a)} Hardware setup showcasing the ASUS WiFi-AP, WARP client and reflector. (b) AoA error and RSSI measured without obfuscation; with nulling; and using \name to delay the path by varying amounts}
    \label{fig:example-result}
\end{figure*}

Specifically, to beamform to the direct path and multipath, we need to have the beamforming vector $S_d(f_i,k)$ and $S_r(f_i,k)$ for the direct path and reflected path respectively~\cite{beam_forming_doc}. The specific beamforming angles are predicted via a downlink channel measurement at the user. Hence, the Wi-Fi user will use the precoding weight of $S_d(f_i,k)e^{-j2\pi f_i d_\mathrm{obf}/c}+S_r(f_i,k)$, where $d_\mathrm{obf}$ indicates the additional path length that Wi-Fi user wants to add to obfuscate its location for Wi-Fi AP as shown in Fig.~\ref{fig:mirage-idea}(d). We ensure that $d_\mathrm{obf} > d_\mathrm{multipath} - d_\mathrm{direct}$ via the same downlink channel measurements. Then, the Wi-Fi AP will receive the following signals:
\begin{align}
\label{eq:beam:delay}
     Y(f_i, k)&=H(f_i,k)(S_d(f_i,k)e^{-j2\pi f_i \frac{d_\mathrm{obf}}{c}} + S_r(f_i,k))X(f_i,k)\nonumber \\
             &=(H_d(f_k, k)+H_r(f_i,k))(S_d(f_i,k)e^{-j2\pi f_i\frac{d_\mathrm{obf}}{c}}\nonumber \\
             & \quad \quad \quad \quad  + S_r(f_i,k))X(f_i,k)\nonumber\\
             &=(H_d(f_k, k)S_d(f_i,k)e^{-j2\pi f_i\frac{d_\mathrm{obf}}{c}} \nonumber\\
             & \quad \quad \quad \quad + H_r(f_i,k)S_r(f_i,k))X(f_i,k)
\end{align}
Beamforming in the directions of the direct path ensures weaker energy directed towards strongest multipath and vice versa. Hence, in Eqn.~\ref{eq:beam:delay}, the cross terms $H_d(f_k, k)S_r(f_i,k)$ and $H_r(f_i,k)S_d(f_i,k)e^{-j2\pi f_i d_\mathrm{obf}/c}$ are ignored. Thus, the wireless channel is $H_d(f_k, k)S_d(f_i,k)e^{-j2\pi f_i d_\mathrm{obf}/c}+H_r(f_i,k)S_r(f_i,k)$, which will only contain direct path with delay of $d_\mathrm{obf}$ and multipath without any delay.

\section{Implementation}\label{sec:results}


\textbf{Wi-Fi AP.} We use ASUS RT-AC86U~\cite{pizarro2021accurate} Wi-Fi AP as the compromised snooper for Wi-Fi user's location. The Wi-Fi AP is instrumented with a uniform linear array of four antennas with spacing of $0.026$ m. The WiFi AP can estimate the uplink channel for indoor localization with SpotFi algorithm. Specifically, Wi-Fi AP will create angle-distance profile via SpotFi for localization by identifying AoA of direct path with the least travelled distance.

\textbf{Wi-Fi User.} We use WARP software defined radio~\cite{warp} as the Wi-Fi user, which is also instrumented with four antennas. The antenna spacing at the Wi-Fi user is 0.026m to achieve accurate nulling and beamforming at the Wi-Fi user. After Wi-Fi user obtains downklink channel, they will `precode' the data stream for location obfuscation with \name as dicussed in Sec.~\ref{sec:mirage-des}. 

\textbf{Setup.} We do experiments in the cluttered indoor environment as shown in Fig.~\ref{fig:example-result}(a). For the proof of concept, there are mainly two paths (i.e., direct path and multipath) between the Wi-Fi AP and user. The multipath gets reflected by the reflector (purple box) shown in the figure. Wi-Fi user (black circle) will communicate with Wi-Fi AP (red box) using 802.11n protocol with bandwidth of 20MHz at the center frequency of 5180MHz. 

\section{Evaluation}
We evaluate \name with the above implementation and demonstrate through real-world experiments and some simulations the capabilities of \name's obfuscation of users AoA and locations.
\subsection{\name's Performance} We first demonstrate that \name will disable the Wi-Fi AP to localize the Wi-Fi user, while \name will not degrade the normal communication throughput between the Wi-Fi AP and user. As shown in Table~\ref{table:aoa_rssi}, with nulling to the direct path approach, SpotFi algorithm will not identify the direct path correctly due to the AoA error of $62^{\circ}$. However, nulling to the direct path will decrease RSSI by 6dBm in comparison to the standard communication with no obfuscation. So, nulling to the direct path will degrade the network throughput. When we employ \name and add different delays to the direct path (i.e., 20m, 30m and 40m), the AoA error becomes significant (i.e., $58^{\circ}$, $61^{\circ}$ and $53^{\circ}$) which will disable SpotFi for accurate localization. Additionally, the RSSIs do not change significantly in comparison to the beamforming without delay and standard communication without \name. 

\begin{figure}
    \centering
    \includegraphics[width=\linewidth]{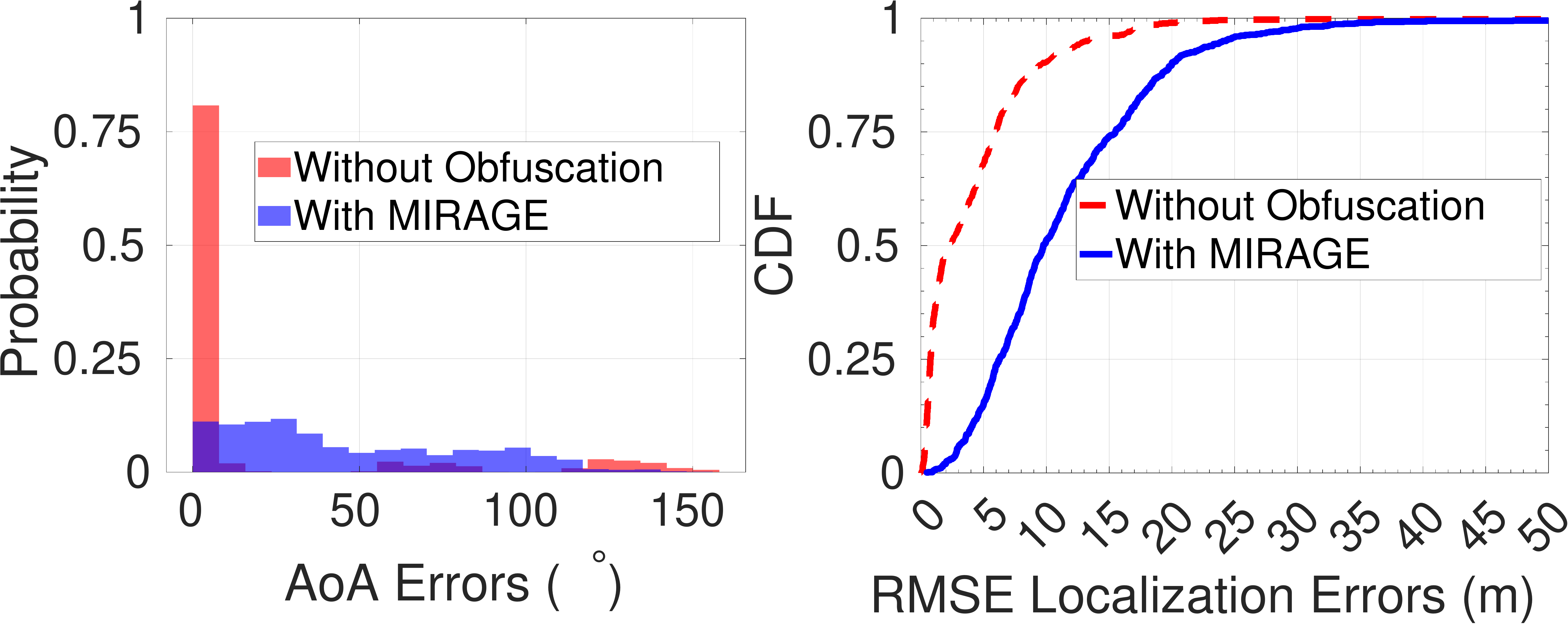}
    \caption{Obfuscation performance: (Left) Absolute AoA prediction errors probabilistic distribution with and without \name. (Right) RMSE Localization CDF with and without \name.}
    \label{fig:location-obf}
\end{figure}



\begin{figure*}[ht]
    \centering
    \includegraphics[width=0.9\linewidth]{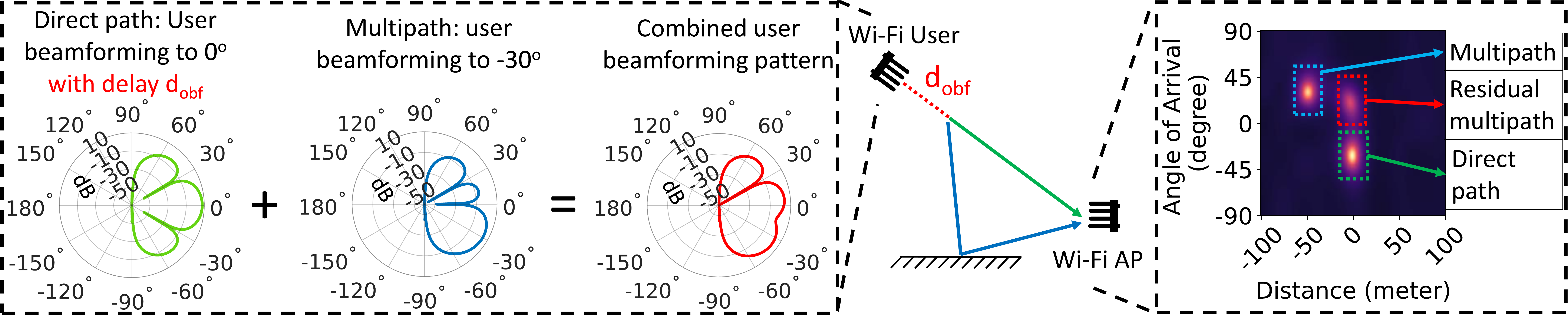}
    \vspace{1pt}
    \caption{Imperfect beamforming introduces residual peak in angle-distance profile (red dotted box) due to the power leakage from the side lobe of the direct path beamforming. \textit{(left inset)} Antenna beamforming patterns transmitted by the user; \textit{right inset} received SpotFi profile at a commodity Wi-Fi AP}
    \label{fig:beam_null_mot}
\end{figure*}

\subsection{Location Obfuscation Performance}
Finally, it is important to understand how much location and angular obfuscation does \name provide for a large deployment scenario. To understand them, we have deployed \name in a simulated environment of $40$~m$\times30$~m, with 4 APs at the center of each wall, where each wall behaves as a perfect reflector and there are additional reflectors placed within the environment to create multipath for significant data points. With this setup we test the obfuscation \name provides to AoA and thus the user location.

\noindent\textbf{AoA accuracy:} We have computed the AoAs of the users moved to 500 random positions within this environment and observed the absolute AoA errors' distribution as shown in Figure~\ref{fig:location-obf}(left). We can see that while the AoA error distributions before Obfuscation had an average error of $20^\circ$. These AoA errors increase $2\times$, as shown in the blue histogram with a mean of $46^\circ$. Demonstrating that \name's AoA obfuscation is successful.

\noindent\textbf{Localization accuracy:} We use the above AoAs to perform both triangulation and AoA+ToF based user localization, and present the errors for both the algorithms together for before (in red) and after (in blue) obfuscation as shown in Figure~\ref{fig:location-obf}(right), where we can see that while before obfuscation the attacker could get the user location with an average precision of $2.3$m and $10$ m after obfuscation the average precision of the user's location known to the attacker, a $5\times$ poorer location accuracy. With an error of this magnitude, the attacker cannot extract any meaningful information either, showing the strengths in \name's AoA obfuscation techniques.








\section{Related Work}\label{sec:related}

The most commonly employed location obfuscation technique is to randomize the user device's MAC address to prevent an attacker from uniquely identifying a user. However, these MAC address randomization approaches can either be easily broken or disabled by the user device~\cite{matte2018spread}. 

Hence, many works look towards disabling an AP's ability to localize users. For signal-strength based techniques, obfuscating the strength of wireless signals is used. Authors in~\cite{zhu2018tu} introduces several approaches (i.e., geofencing with electromagnetic shielding paint on the walls and jamming with extra signal generators), however this interrupts ongoing wireless communication between the Wi-Fi AP and user. These techniques hence comprise a weak defense against location snooping. On the other hand, Fine Timing Measurement (FTM)~\cite{ibrahim2018verification} based localization, introduced in 802.11 standard, considered to be secure~\cite{schepers2022privacy,ftm_secure}, can also be leveraged for privacy invasive localization~\cite{wipeep}. 

Considering the poor defenses against signal-strength or FTM based localization techniques, some prior works look towards modifying the wireless sensing environment. However, these systems deploy additional hardware in the environment affecting widescale adoption. For example, PhyCloak~\cite{qiao2016phycloak} requires a full duplex radio co-located with the user for location obfuscation; IRShield~\cite{staat2022irshield} requires to deploy a smart surface in the wireless sensing environment to distort the wireless channel for location obfuscation. Additionally, PhyCloak and IRSheild may interrupt user communication. RF-Protect~\cite{shenoy2022rf} deploys a reflector in the environment to obfuscate the wideband FMCW signals; Aegis~\cite{yao2018aegis} deploys an additional radio instrumented with amplifier and antennas to obfuscate the wireless signals. 

To overcome the limitations of prior work, \name simply uses the beamforming capability of end user devices (e.g., smartphone) to obfuscate the user location without affecting the ongoing wireless communication. 
\section{Discussion and Future Work}\label{sec:conlusion}

In this section, we discuss the implications and limitations of \name's current implementation and propose the future research direction for privacy-preserving wireless sensing.

\noindent\textbf{{Improving User Location Obfuscation: }}
We have demonstrated \name's obfuscation for the Wi-Fi user's location through beamforming and delaying through a few examples. However in \name's design we make a critical assumption. We ignore the cross terms in Eqn.~\ref{eq:beam:delay} when adding delay to the direct path due to weaker energy directed towards the reflected path. However, in some cases where there are strong reflection in the environment, these cross terms cannot be ignored. An exemplary case is shown in Fig.~\ref{fig:beam_null_mot}. In this scenario, we beamform towards $0^\circ$ and add a delay of $d_\mathrm{obf} = 40 m$ to the direct path (green pattern) and simultaneously beamform towards the multipath at $-30^\circ$ (blue pattern). This creates a combined beamforming pattern (red) shown in the left inset. However, due to some side-lobe leakage in the direction of multipath and the presence of a strong reflector, we inadvertently add delay to the multipath as well. This creates a residual peak (in red dotted box) along with the delayed direct path peak (green box). This abnormal angle-distance profile will expose \name's spoofing and an attacker can potentially extract the correct direct path. 

\noindent\textbf{Multiple Collaborative Wi-Fi APs: } 
Currently, \name only considers one Wi-Fi AP deployed in the environment to steal the Wi-Fi user's location information. In the enterprise network, there are multiple Wi-Fi APs deployed in the building to achieve larger coverage and better service. These Wi-Fi APs can collaborate with each other to localize the Wi-Fi user, which will significantly improve the localization accuracy. To defend against multiple APs, the Wi-Fi user needs to obfuscate AoA information extracted by each of them, which will require the Wi-Fi user to create more fine-grained beams with more antennas and eliminate the interference across different beams. 

\noindent\textbf{Dynamic Wireless Environment: } 
To accurately obfuscate AoA information, the Wi-Fi user needs to accurately estimate the wireless channel either with CSI feedback from the Wi-Fi AP which will require the collaboration of the attacker (i.e., Wi-Fi AP) and introduce the overhead due to the downlink communication of feedback, or leveraging the property of wireless channel reciprocity. In another perspective, the dynamic environment will decrease the accuracy of wireless localization due to the dynamic clutters in the environment, which will require the frequent channel estimation. 
 
\noindent\textbf{Wi-Fi User's Beamforming/Nulling Capability: } 
The performance of our obfuscation depends on Wi-Fi user's beamforming/nulling capability. More number of antennas at the Wi-Fi user, more accurate estimation of AoA. This is because the Wi-Fi user can always shine the narrow beam towards the Wi-Fi user for the purpose of obfuscation without interfering the reflected path. The commodity smartphones are usually instrumented with three or more antennas~\cite{antenna_phone}, which are enough for our user location obfuscation. For example, Iphone 13 supports 4x4 MIMO for 5G and 2x2 MIMO for Wi-Fi 6 (802.11ax), which will enable them to leverage \name for location obfuscation.

\noindent\textbf{How Much Location Obfuscation is Needed?: }
\name is designed to defend against the fine-grained wireless localization algorithms (e.g., Spotfi~\cite{spotfi}), which can provide the localization error of decimeter. Therefore, any location obfuscation that can make the localization error more than decimeter will disable the Wi-Fi AP to accurately localize the Wi-Fi user and derive the context sensing information from the user's location. However,  the Wi-Fi AP may just employ the coarse-grained localization algorithms (e.g., RSSI-based indoor localization~\cite{radar}) to localize the Wi-Fi user for context sensing (e.g., to know if people are at home or not). In this case, Wi-Fi user can leverage the nulling technique illustrated in Section~\ref{sub:nulling} to hide his/her location or beamform to a far distance, while it will degrade the network throughput as the signal strength will be significantly degraded.

\bibliographystyle{abbrv}
\bibliography{reference}

\begin{thebibliography}{10}

\bibitem{wipeep}
A.~Abedi and D.~Vasisht.
\newblock Non-cooperative wi-fi localization \& its privacy implications.
\newblock In {\em Proceedings of the 28th Annual International Conference On
  Mobile Computing And Networking}, pages 126--138. ACM, 2022.

\bibitem{alliance20155g}
N.~Alliance.
\newblock 5g white paper.
\newblock {\em Next generation mobile networks, white paper}, 1(2015), 2015.

\bibitem{radar}
V.~Bahl and V.~Padmanabhan.
\newblock {RADAR: An In-Building RF-based User Location and Tracking System}.
\newblock INFOCOM, 2000.

\bibitem{ibrahim2018verification}
M.~Ibrahim, H.~Liu, M.~Jawahar, V.~Nguyen, M.~Gruteser, R.~Howard, B.~Yu, and
  F.~Bai.
\newblock Verification: Accuracy evaluation of wifi fine time measurements on
  an open platform.
\newblock In {\em Proceedings of the 24th Annual International Conference on
  Mobile Computing and Networking}, pages 417--427, 2018.

\bibitem{spotfi}
M.~Kotaru, K.~Joshi, D.~Bharadia, and S.~Katti.
\newblock {SpotFi: Decimeter Level Localization Using Wi-Fi}.
\newblock SIGCOMM, 2015.

\bibitem{li2005null}
M.~Li and Y.~Lu.
\newblock Null-steering beamspace transformation design for robust data
  reduction.
\newblock In {\em 2005 13th European Signal Processing Conference}, pages 1--4.
  IEEE, 2005.

\bibitem{martin2020ranging}
I.~Martin-Escalona and E.~Zola.
\newblock Ranging estimation error in wifi devices running ieee 802.11 mc.
\newblock In {\em GLOBECOM 2020-2020 IEEE Global Communications Conference},
  pages 1--7. IEEE, 2020.

\bibitem{matte2018spread}
C.~Matte and M.~Cunche.
\newblock {\em Spread of MAC address randomization studied using locally
  administered MAC addresses use historic}.
\newblock PhD thesis, Inria Grenoble Rh{\^o}ne-Alpes, 2018.

\bibitem{beam_forming_doc}
{Natalia Schmid}.
\newblock {beamforming weight design }.
\newblock https://safe.nrao.edu/wiki/pub/Beamformer/WebHome.

\bibitem{pizarro2021accurate}
A.~B. Pizarro, J.~P. Beltr{\'a}n, M.~Cominelli, F.~Gringoli, and J.~Widmer.
\newblock Accurate ubiquitous localization with off-the-shelf ieee 802.11 ac
  devices.
\newblock In {\em Proceedings of the 19th Annual International Conference on
  Mobile Systems, Applications, and Services}, pages 241--254, 2021.

\bibitem{qiao2016phycloak}
Y.~Qiao, O.~Zhang, W.~Zhou, K.~Srinivasan, and A.~Arora.
\newblock $\{$PhyCloak$\}$: Obfuscating sensing from communication signals.
\newblock In {\em 13th USENIX Symposium on Networked Systems Design and
  Implementation (NSDI 16)}, pages 685--699, 2016.

\bibitem{antenna_phone}
{Qualcomm}.
\newblock {Iphone antennas }.
\newblock https://discussions.apple.com/thread/1334188.

\bibitem{qualcomm}
{Qualcomm}.
\newblock {Qualcomm Enterprise Network }.
\newblock
  https://www.qualcomm.com/products/application/wireless-networks/wi-fi-networks/networking-pro-series.

\bibitem{ftm_secure}
{Qualcomm and Android FTM MAC address randomizatoin}.
\newblock {CVE-2020-11287 Detail }.
\newblock https://nvd.nist.gov/vuln/detail/CVE-2020-11287.

\bibitem{warp}
{rice university}.
\newblock {WARP software defined radio }.
\newblock https://warpproject.org/trac/wiki/about.

\bibitem{schepers2022privacy}
D.~Schepers and A.~Ranganathan.
\newblock Privacy-preserving positioning in wi-fi fine timing measurement.
\newblock {\em Proceedings on Privacy Enhancing Technologies},
  2022(2):325--343, 2022.

\bibitem{shenoy2022rf}
J.~Shenoy, Z.~Liu, B.~Tao, Z.~Kabelac, and D.~Vasisht.
\newblock Rf-protect: privacy against device-free human tracking.
\newblock In {\em Proceedings of the ACM SIGCOMM 2022 Conference}, pages
  588--600, 2022.

\bibitem{malls_loc}
I.~Solomiia~Ryfiak.
\newblock {Indoor Positioning Technologies as a Rising Force in Retail Sales}.
\newblock
  \url{https://intellias.com/indoor-positioning-technologies-as-a-rising-force-in-retail-sales/}.

\bibitem{staat2022irshield}
P.~Staat, S.~Mulzer, S.~Roth, V.~Moonsamy, M.~Heinrichs, R.~Kronberger,
  A.~Sezgin, and C.~Paar.
\newblock Irshield: A countermeasure against adversarial physical-layer
  wireless sensing.
\newblock In {\em 2022 IEEE Symposium on Security and Privacy (SP)}, pages
  1705--1721. IEEE, 2022.

\bibitem{chronos}
D.~Vasisht, S.~Kumar, and D.~Katabi.
\newblock {Decimeter-Level Localization with a Single Wi-Fi Access Point}.
\newblock NSDI, 2016.

\bibitem{yao2018aegis}
Y.~Yao, Y.~Li, X.~Liu, Z.~Chi, W.~Wang, T.~Xie, and T.~Zhu.
\newblock Aegis: An interference-negligible rf sensing shield.
\newblock In {\em IEEE INFOCOM 2018-IEEE conference on computer
  communications}, pages 1718--1726. IEEE, 2018.

\bibitem{zhu2018tu}
Y.~Zhu, Z.~Xiao, Y.~Chen, Z.~Li, M.~Liu, B.~Y. Zhao, and H.~Zheng.
\newblock Et tu alexa? when commodity wifi devices turn into adversarial motion
  sensors.
\newblock {\em arXiv preprint arXiv:1810.10109}, 2018.

\end{thebibliography}

\end{document}